\documentclass[doublecol,figures]{epl2} 
\usepackage{graphicx}
\usepackage{amsmath,amssymb}
\usepackage{bm}

\title{Classical Langevin dynamics of a charged particle moving on a sphere 
and diamagnetism: A surprise} 

\shorttitle{Classical diamagnetism}

\author{N. Kumar \inst{1,3}\thanks{email: \email{nkumar@rri.res.in}} 
		\and K. Vijay Kumar \inst{2}\thanks{email: \email{vijayk@physics.iisc.ernet.in}}}
\shortauthor{N. Kumar \etal}

\institute{                    
  \inst{1} Raman Research Institute, Bangalore, India 560080\\
  \inst{2} CCMT, Dept. of Physics, Indian Institute of Science, Bangalore, 
						India 560012 \\
	\inst{3} Jawaharlal Nehru Centre for Advanced Scientific Research, 
					Bangalore, India 560064
}

\abstract{It is generally known that the orbital diamagnetism of a classical
system of charged particles in thermal equilibrium is identically zero -- the
Bohr-van Leeuwen theorem. Physically, this null result derives from the exact
cancellation of the orbital diamagnetic moment associated with the complete
cyclotron orbits of the charged particles by the paramagnetic moment subtended
by the incomplete orbits skipping the boundary in the opposite sense. Motivated
by this crucial, but subtle role of the boundary, we have simulated here the
case of a finite but \emph{unbounded} system, namely that of a charged particle
moving on the surface of a sphere in the presence of an externally applied uniform 
magnetic field. 
Following a real space-time approach based on the classical Langevin equation, we have 
computed the orbital magnetic moment which now indeed turns out to be non-zero, and has 
the diamagnetic sign. To the best of our knowledge, this is the first report of 
the possibility of finite classical diamagnetism in principle, and it is due to the avoided 
cancellation.}

\begin{document}

\maketitle


In this work we re-visit the problem of the absence of classical diamagnetism of
a system of charged particles in thermal equilibrium. This vanishing of the
classical diamagnetism in equilibrium is generally referred to as the Bohr-van
Leeuwen theorem \cite{vanLeeuwen,BohrThesis,vanVleck,Peierls}.  The fact that
classically the orbital diamagnetic moment vanishes is quite contrary to our
physical expectations inasmuch as a charged particle (of charge $-e$, position
${\bf r}(t)$, and velocity ${\bf v}(t)$ at time $t$), say, orbiting in a plane
perpendicular to the magnetic field ${\bf B}$ under its Lorentz force should
have an orbital magnetic moment ${\bf M} \, ( =-e/2c \big[ {\bf r}(t) \times
{\bf v}(t) \big] )$, where $c$ is the speed of light, and with a diamagnetic
sign as dictated by the Lenz's law (see e.g. \cite{Jackson}).  Formally, 
the vanishing of the classical diamagnetic moment follows from the well known
fact that the canonical partition function involves the Hamiltonian for the 
charged particle (coupled minimally to the static magnetic field) and a 
simple shift of the canonical momentum variable in the integration makes the
partition function field-independent, giving zero magnetic moment 
\cite{vanVleck}.
 
Physically, the vanishing of the classical diamagnetism is due, however, to a subtle role
played by the boundary of the finite sample \cite{vanLeeuwen,BohrThesis,vanVleck,Peierls}. It turns out that the diamagnetic
contribution of the completed cyclotron orbits of the charged particles orbiting
around the magnetic field in a plane perpendicular to it, is cancelled by the
paramagnetic contribution of the incomplete orbits skipping the boundary in the
opposite sense in a cuspidal manner.  The cancellation is exact, and that is the
surprise. This cancellation was demonstrated explicitly some time back
\cite{KumarJayannavar} for the case of a harmonic-potential ($V(r)=k r^2 / 2$)
confinement, which is equivalent to a soft boundary, and finally letting the
spring constant $k$ go to zero.  The treatment was based on the classical
Langevin equation \cite{Coffey}, and the magnetic moment  ${\bf M} =-e/2c \big[
{\bf r}(t) \times {\bf v}(t) \big]$ was calculated in the infinte-time limit --
the Einsteinian approach to statistical mechanics. The ordering of the two
limits namely $k \to 0$ (the deconfinement limit) and $t \to \infty$ (the
infinite time limit), however, turned out to be crucial and physically
meaningful -- one must let $t \to \infty$ first and then let $k \to 0$.  This
ensures that the particle is affected by the boundary, or the confinement. Thus
one had to conclude that any orbital diamagnetism observed in an experiment is
essentially of quantum-mechanical origin, as indeed was derived first by Landau
\cite{Landau}. In the quantum case, the above cancelllation of the bulk and the
boundary contributions turns out to be incomplete. But again, the order of the
two limits is all important and was implicit in the treatment of Landau
\cite{vanVleck}. It was shown more explicitly by Darwin \cite{Darwin}. In fact,
one could just use the quantum Langevin equation \cite{Zoller} and derive
essentially the Landau result  by properly taking the above `Darwin limit'. The
calculated diamagnetic moment is, however, found to depend on the frictional term occuring
in the quantum Langevin equation \cite{DattaguptaPRL,DattaguptaKumarJayannavar}.

The subtle but essential role of the boundary in all these treatments has
motivated us to examine the diamagnetism for a classical system that has \emph{no
geometrical boundary} -- \emph{a finite unbounded} system such as a charged
particle moving on the surface of a sphere under the appropriate Langevin
dynamics in the presence of a uniform external magnetic field. We were pleasantly
surprised to find that the numerically computed orbital magnetic moment turned
out to be non-zero, and indeed diamagnetic. To the best of our knowledge, this is
the first example reported on non-zero orbital diamagnetism in principle for a classical
system. It arises explicitly from avoided cancellation as the system has no
boundary.


Consider a charged particle (charge -$e$ and mass $m$, an electron say) moving
on the surface of a sphere of radius `$a$', in the presence of a uniform
externally applied magnetic field ${\bf B}$ directed along the $z-$axis. The
particle motion is described by the following classical Langevin equation
\begin{equation} m \frac{d {\bf v}}{dt} = -\frac{e}{c} ({\bf v} \times {\bf B})
- \Gamma {\bf v} + \sqrt{2 \Gamma k_B T} \, {\bf f}(t) \end{equation} where
$\Gamma$ is the friction coefficient, $k_BT$
the thermal energy, and ${\bf f}$ is a zero-mean $\delta-$correlated Gaussian
random noise, i.e., $\langle f_{\alpha}(t) f_{\beta}(t') \rangle =
\delta_{\alpha\beta} \delta(t-t')$.  We recall here that in this real space-time
(Einsteinian) approach to statistical mechanics, the long-time limit
($t\to\infty$) of the above stochastic evolution is expected to give the thermal-equilibrium
properties. Note that there is no modification of the dissipation ($\Gamma$) 
and the related noise term (${\bf f}(t)$) due to the magnetic field \cite{Coffey}.

Specializing now to the spherical-polar coordinates appropriate to the motion on
the surface of the sphere ($r=a, \theta, \phi$), the Langevin equation reduces
to 
\begin{eqnarray} a\Big[ \frac{d^2 \theta}{dt^2} - \sin\theta
\cos\theta \, \big( \frac{d \phi}{dt} \big)^2 \Big] \, \hat{\bm \theta} 
\nonumber \\
&& \hspace{-5cm}
+ a \Big[ \sin\theta \, \frac{d^2 \phi}{dt^2} + 2 \, \cos\theta \, 
	\frac{d \theta}{dt} \, \frac{d \phi}{dt} \Big] \, \hat{\bm \phi} 
	\nonumber \\
	&& \hspace{-4cm}
	= - \frac{eB}{mc} \,
	a \, \Big[ \frac{d \theta}{dt} \,  \hat{\bm \theta} 
	+ \sin\theta \, \frac{d \phi}{dt} \, \hat{\bm \phi} \Big] \times 
	(\hat{\bf r} \, \cos\theta) 
\nonumber
\\ 
&&  \hspace{-3cm} 
- \frac{a \Gamma}{m} \Big[\frac{d \theta}{dt} \,  \hat{\bm
\theta} + \sin\theta \, \frac{d \phi}{dt} \, \hat{\bm \phi} \Big] 
\nonumber
\\
&&  \hspace{-3cm} 
+
\frac{\sqrt{2 \Gamma k_B T}}{m} \Big( f_{\theta} \, \hat{\bm \theta} + f_{\phi}
\, \hat{\bm \phi} \Big), \label{eq:LangevinEquationSpherical} \end{eqnarray}
where ${\bf \hat{r}}$, $\hat{\bm \theta}$ and $\mathbf{\hat{\bm
\phi}}$ are the unit vectors directed along the radial ($r$), polar ($\theta$)
and the azimuthal ($\phi$) directions. Also, $f_{\theta}$ and $f_{\phi}$ are the
forcing noise terms acting along the $\theta$ and the $\phi$ directions
respectively. More conveniently, we re-write equation
(\ref{eq:LangevinEquationSpherical}) in the dimensionless form
\begin{eqnarray} \ddot{\theta} -
\sin\theta \cos\theta \dot{\phi}^2 &=& -\frac{\omega_{c}}{\gamma}  \sin\theta
\cos\theta \dot{\phi} - \dot{\theta} + \sqrt{\eta} f_{\theta}, \\ \sin\theta
\ddot{\phi} + 2 \cos\theta \dot{\theta} \dot{\phi} &=& \frac{\omega_{c}}{\gamma}
\cos\theta \dot{\theta} - \sin\theta \dot{\phi} + \sqrt{\eta} f_{\phi},
\label{eq:eqn_of_motion}
\end{eqnarray} where we have introduced the cyclotron
frequency $\omega_{c} = eB/mc$, the frictional velocity relaxation rate
$\gamma=\Gamma/m$, the thermal forcing strength $\eta=2 k_B T/ (m a^2
\gamma^2)$, and the dimensionless time $\tau=\gamma t$.  Note that $\eta$ is
also a dimensionless quantity.  Here overhead dots denote differentiation with
respect to the dimensionless time $\tau$. The physical quantity of interest is
the ensemble averaged orbital magnetic moment \begin{equation} \langle M(\tau)
\rangle = -\frac{e}{2 c} \, \gamma \, a^2 \langle \sin^2\theta(\tau)
\dot{\phi}(\tau) \rangle \end{equation} in the long-time limit, where $\langle
\cdots \rangle$ denotes the ensemble average over the different realizations of
the stochastic forces $f_{\theta}$ and $f_{\phi}$.
 
We now rewrite the above second-order differential Langevin equations
(\ref{eq:eqn_of_motion}) as four coupled first-order equations for $\theta$, $x
\, (\equiv \dot{\theta})$, $\phi$, and $y \, (\equiv \dot{\phi})$, which are
then solved numerically using a simple Euler-Maruyama scheme
\cite{KloedenPlatenSchurz} with a time-step $\Delta \tau = 10^{-2}$. Averages
are evaluated over $n=10^{6}$ noise realizations. The number of realizations,
though quite large, is necessarily finite, and so we resort to double average
$\langle \langle \cdots \rangle \rangle$ denoting averaging over the ensemble as
well as over time. This gives for the equilibrium magnetic moment
\begin{equation} M_{eq} = \langle \langle M(\tau) \rangle \rangle \equiv
\frac{1}{\tau_{max}} \, \int_{0}^{\tau_{max}} \, \langle M(\tau) \rangle \,
d\tau \label{eq:meq} \end{equation} as $\tau_{max} \to \infty$.  In the context
of numerical simulation, we have to be careful at the singular polar points
$\theta=0$ and $\theta=\pi$, where $1/\sin\theta$ diverges. This is regularized
by replacing $\sin\theta$ by $\sqrt{\sin^2\theta + \epsilon}$ where $\epsilon$
is a small positive quantity taken to be of order $\Delta \tau$. Further,
inasmuch as the physical motion is restricted to $0\leq\theta<\pi$ and $0\leq
\phi < 2 \pi$, while mathematically, however, the equations
(\ref{eq:eqn_of_motion}) can evolve outside these bounds, we have to set in our
numerical simulation the following conditions: If $\theta(\tau) < 0$, then
$\theta(\tau) \to -\theta(\tau)$, $x(\tau) \to -x(\tau)$, $\phi(\tau) \to
\phi(\tau-\Delta \tau) + \pi$; and if  $\theta(\tau) > \pi$, then $\theta(\tau)
\to 2\pi-\theta(\tau)$, $x(\tau) \to -x(\tau)$, $\phi(\tau) \to \phi(\tau-\Delta
\tau) - \pi$.  This takes care of the trajectories that happen to pass through
the poles.  The choice of initial conditions on $\theta$ and $\phi$, and their
time derivatives, turns out to be irrelevant for the long-time ensemble averaged
behavior as indeed is validated by our numerical simulation.

\begin{figure} 
\onefigure[width=0.8\columnwidth]{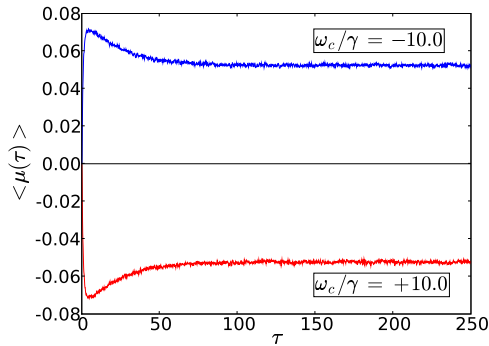}
\caption{{\small (Color online)} Plot of the ensemble
averaged dimensionless magnetic moment $\mu(\tau)$ as a function of the
dimensionlesss time $\tau$ for $\omega_{c}/\gamma=\pm 10.0$ and $\eta=1.0$.
Clearly the moment can be seen to be odd in the magnetic field ${\bf B}$ and is
diamagnetic.}
\label{fig:mu_vs_tau}
\end{figure}

In Fig. \ref{fig:mu_vs_tau}, we have plotted the dimensionless magnetic moment
$\langle \mu(\tau) \rangle = 2c/(e \gamma a^2) \, \langle M(\tau) \rangle$ as a
function of $\tau$ for certain choice of $\omega_{c}/\gamma$ and $\eta$.  As can
be readily seen, the moment is diamagnetic and odd in the magnetic field. Also,
it can shown to be independent of the sign of the charge (electron or hole) as
indeed it must be. The fluctuations seen in the figure are statistical
fluctuations due to the finiteness of the number of realizations used for
ensemble averaging. These are thus statistical fluctuations -- these will, and
indeed do, decrease with increasing number of noise realizations $n$.

\begin{figure}
\onefigure[width=\columnwidth]{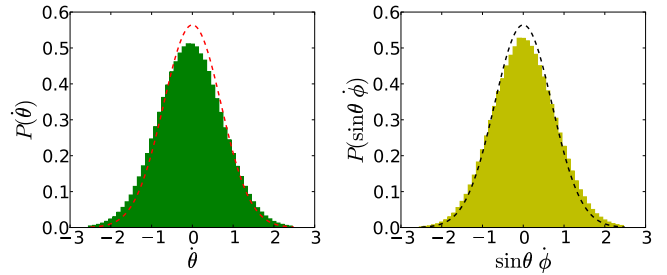}
\caption{{\small (Color online)} The velocity distribution 
on a sphere in the presence of magnetic field computed from the Langevin dynamics 
simulation. The dashed curve is the corresponding Maxwellian distribution. Here $\omega_{c}/\gamma = 10.0$ and $\eta = 1.0$.} 
\label{fig:vel-dist}
\end{figure}

For completeness, we have also plotted the computed velocity distributions (the $\theta$ 
and $\phi$ components) in fig. \ref{fig:vel-dist} and these are seen to be 
essentially Maxwellian as expected, with the correct mean square values consistent with 
the fluctuation-dissipation theorem. In fact there is no dependence on the external 
magnetic field.

\begin{figure} 
\onefigure[width=0.8\columnwidth]{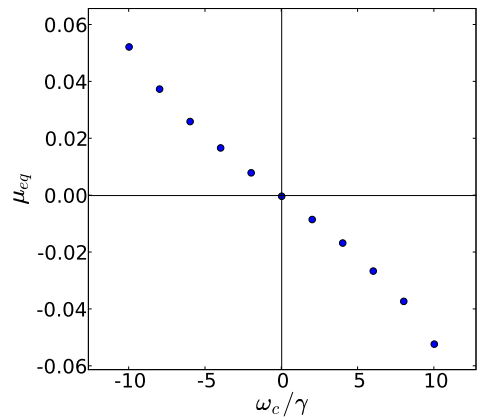}
\caption{{\small (Color online)} Plot of the dimensionless
magnetic moment $\mu_{eq}$ as a function of $\omega_{c}/\gamma$ (proportional to
the magnetic field) for $\eta=1.0$. Again the moment is found to be odd in ${\bf
B}$ and is diamagnetic in sign.} 
\label{fig:m_vs_b}
\end{figure}

Fig. \ref{fig:m_vs_b} shows the variation of the dimensionless magnetic moment
$\mu_{eq}$ (corresponding to $M_{eq}$) with the magnetic field $\omega_c /
\gamma $ (which is proportional to ${\bf B}$). The plot shows an essential
linear response which is  diamagnetic.

\begin{figure} 
\onefigure[width=0.8\columnwidth]{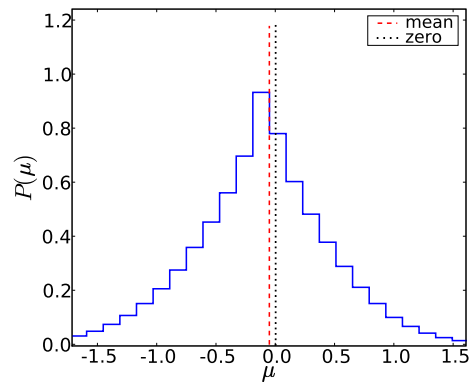}
\caption{{\small (Color online)} Plot of the long-time
probability density $P(\mu)$ against $\mu$ giving the ensemble fluctuations
about $\mu_{eq}$. The latter is clearly non-zero and diamagnetic. The
fluctuations are seen to be large compared to the mean value.  Here
$\omega_{c}/\gamma = 10.0$ and $\eta = 1.0$.} 
\label{fig:distribution}
\end{figure}

In Fig. \ref{fig:distribution}, we have plotted the probability density
$P(\mu)$ of the statistical mechanical fluctuations about the
equilibrium value $\mu_{eq}$. The distribution for the chosen values of the
parameters is quite broad relative to the mean. (The corresponding plot for a
system \emph{with} a boundary is indeed known to be broad
\cite{DibyenduKumar}. Of course, in that case the mean is zero.)

We now return to the main point of this puzzle, namely that the classical Langevin 
dynamics for this finite unbounded system gives a non-zero diamagnetic moment, and yet
a straightforward calculation using the canonical partition function with a minimally coupled 
Hamiltonian gives a free energy which is independent of the magnetic field, and therefore, a 
zero field-derivative of the latter implying zero diamagnetism. Now, the classical Langevin 
dynamics provides a real space-time picture of the charged particle motion under the 
influence of fluctuations and the concomitant dissipation. 
Its long-time limit is expected to describe thermal equilibrium.
And sure enough it does reproduce 
the Maxwellian velocity distribution (fig. \ref{fig:vel-dist}). 
Moreover, it is manifestly gauge-invariant because 
it involves the magnetic field directly without invoking a vector potential (indeed, in 
classical electrodynamics, the vector potential is essentially a matter of convenience, 
unlike in the case of quantum mechanics). 
Also, the computed diamagnetism is consistent with the Lenz's law. 
On the other hand, the canonical treatment based on the Hamiltonian underlies all of 
classical statistical mechanics, but it gives zero classical diamagnetism. 
Our resolution of this puzzle is as follows. It is known that classically the static 
magnetic field does no work on the moving charge inasmuch as the Lorentz force 
($-\frac{e}{c} ({\bf v} \times {\bf B})$) acts perpendicular 
to the instantaneous velocity vector. However, such a gyroscopic force can still alter the motion of the charged particle so as to give a non-zero magnetic moment \emph{without changing 
its energy}. In fact, it induces a correlation between the velocity and the transverse 
acceleration due to the Lorentz force. Clearly, such a subtle dynamical correlation, without change of energy, is not captured by the equilibrium partition function. But, the Langevin dynamics manifestly treats this gyroscopic Lorentz force through the equation of motion. 
It is thus our view that the real space-time treatment based on Langevin dynamics takes into 
account these subtle correlations involving velocity and the transverse acceleration 
caused by the Lorentz force without changing the energy, which is missing from the 
usual partition function.  Of course this disagreement is only for the special case of a strictly unbounded classical system in an external magnetic field.

Thus, we are forced to admit that there are the following two alternatives -- either 
there is indeed non-zero classical diamagnetism for an unbounded finite system as 
under consideration, or the classical Langevin dynamics fails to describe in 
the long-time limit the thermal equilibrium as described by the classical partition 
function in the presence of a magnetic field \cite{SameHamiltonian}. Either way, we have a 
non-trivial result.

Finally, it will be apt at this stage to make a few comments, bearing on the physical 
realizability of such a classical system. 

First, we recall that the
non-zero diamagnetic moment for the classical system discussed above is due
entirely to the absence of a boundary -- the avoided cancellation for a finite
but \emph{unbounded} system. Now, for the case of the quantum mechanical
(Landau) diamagnetism too there is a cancellation, but it is incomplete
\cite{vanVleck}.  Hence the smallness of the Landau diamagnetism in general. We
may reasonably expect then that the quantum mechanical diamagnetism for a finite
but \emph{unbounded} system too should be different, probably larger, because of the avoided
cancellation \cite{WorkInProgress}.  Second, our classical treatment is 
valid in principle for a finite but unbounded system, i.e., a \emph{strictly} closed
two-dimensional surface (the charged particle moving on the surface of a sphere). It may, 
however, be physically realized to an approximation. Thus, we could consider a
dielectric microsphere coated with an ultrathin layer of a conducting material
having small carrier concentration at room temperature, e.g., a non-degenerate
system with the degeneracy temperature much smaller than the room temperature,
as in the case of a doped high-mobility semiconductor. (By ultrathin we mean
here a thickness $\ll$ the thermal de Broglie wavelength of charge carriers so
as to \emph{freeze out the radial motion quantum mechanically}, making the system 
essentially a
two-dimensional classical gas of charged particles moving on the surface of the sphere. 
Thus, quantum mechanics helps us realize a dimensional reduction -- an essentially two-dimensional closed surface). We could then consider a
finite volume fraction of an inert medium (paraffin say) occupied by the above
microspheres.  This system should have a measurable diamagnetic response which will be 
essentially classical. 
Third, it should be interesting to consider more general geometries such as that of a
triaxial ellipsoid where different axes ratios can mimick very different
physical situations. Perhaps much more interesting will be to try out topologies
other than that of a sphere and look for qualitative differences
\cite{WorkInProgress}. The numerically estimated value of the the diamagnetic moment
for a charged particle (say, electron) moving on a sphere of radius $a=100 \mu m$
for $B \simeq 5$ kilogauss, $\gamma \sim 10^{9} s^{-1}$ turns out to be $\sim 1$
Bohr magneton per electron which is quite large.
Thus we may have a giant classical diamagnetism. Of
course, the measured bulk susceptibility for the physical classical system
suggested above may have much smaller values because of the realizable parameter values. 
But, the point of principle at issue will have been made. It is our hope that 
experimentalists may take note of this possibility.

To conclude, we have shown that a classical system of charged particles moving
on a finite but \emph{unbounded} surface (of a sphere), as described by Langevin dynamics,  
has a non-zero orbital diamagnetic moment. This moment can be large.

\begin{thebibliography}{10}
\bibitem{SameHamiltonian}
This can be readily verified by making the well known shift of the canonical
  momentum variable in the partition function. This also works for the case of
  a particle moving on the surface of a sphere.
\end{thebibliography}

\end{document}